\documentclass{aa}
\input{psfig.sty}
\usepackage{lscape}
\usepackage{graphicx,times}

\begin{document}
\authorrunning{K. C. Steenbrugge et al.}
\titlerunning{LETGS and XMM-Newton observations of NGC~4593}
\title{Chandra LETGS and XMM-Newton observations of NGC~4593}
\subtitle{}

\author{K.C. Steenbrugge\inst{1}, J.S. Kaastra\inst{1}, A. J. Blustin\inst{2}, G. Branduardi-Raymont\inst{2}, M. Sako\inst{3}\thanks{{\it Chandra} Postdoctoral Fellow}, E. Behar\inst{4}, S. M. Kahn\inst{5}, F. B. S. Paerels\inst{5}, R. Walter\inst{6}}
\offprints{K.C. Steenbrugge}
\mail{K.C.Steenbrugge@sron.nl}
\institute{ SRON National Institute for Space Research, Sorbonnelaan 2, 3584 CA Utrecht, The Netherlands
\and Mullard Space Science Laboratory, University College London, Holmbury St Mary, Dorking Surrey RH5 6NT, UK 
\and Theoretical Astrophysics and Space Radiation Laboratory, California Institute of Technology, MC 130-133, Pasadena, CA  91125, USA
\and Physics Department, Technion 32000, Israel
\and Columbia Astrophysics Laboratory, Columbia University, 550 West 120th Street, New York, NY 10027, USA 
\and University of Geneva, Geneva, Switzerland}
\date{Received  / Accepted  }

\abstract{In this paper, we analyze spectra of the Seyfert 1 galaxy NGC~4593 obtained with the {\it Chandra} Low Energy Transmission Grating Spectrometer (LETGS), the Reflection Grating Spectrometer (RGS) and the European Photon Imaging Camera's (EPIC) onboard of XMM-{\it Newton}. The two observations were separated by $\sim$7 months. In the LETGS spectrum we detect a highly ionized warm absorber corresponding to an ionization state of 400$\times$10$^{-9}$ W m, visible as a depression at $10-18$ \AA. This depression is formed by multiple weak Fe and Ne lines. A much smaller column density was found for the lowly ionized warm absorber, corresponding to $\xi$ = 3$\times$10$^{-9}$ W m. However, an intermediate ionization warm absorber is not detected. For the RGS data the ionization state is hard to constrain. The EPIC results show a narrow Fe K$\alpha$ line. 
\keywords{AGN: Seyfert 1 --
X-ray: spectroscopy --
AGN: individual: NGC~4593}}

\maketitle

\section{Introduction}
\label{sec:intro}
NGC~4593 is classified as a Seyfert 1 galaxy (Simkin, Su \& Schwarz \cite{simkin}). In the optical it has two prominent spiral arms and a clear central bar (Santos-Lle\'o et al. \cite{santos94}).  Because of its low redshift of 0.0084 (Paturel et al. \cite{paturel}), it is one of the X-ray brightest Seyfert galaxies in the sky. A further advantage is the low galactic column density of 1.97 $\times$ 10$^{24}$ m$^{-2}$ (Elvis, Wilkes \& Lockman \cite{elvis}). As a result it has been intensively studied in the X-ray and UV bands.
\par
A strong absorption-like feature in the spectrum between $10-18$ \AA~($0.7-1.2$ keV) was detected in the ASCA and {\it Beppo}SAX observations (George et al. \cite{george}, Guainazzi et al. \cite{guai}). Therefore, it was inferred that there were strong K-shell absorption edges for \ion{O}{vii} and \ion{O}{viii} at 0.74 and 0.87 keV, respectively. From the depth of the edges the optical depth was derived as~0.3 and 0.1 for the \ion{O}{vii} and \ion{O}{viii} edge, respectively (Reynolds \cite{reynolds}). It was thus deduced that NGC~4593 has a strong warm absorber, which would result in a complex absorption line spectrum, if observed with the current high resolution observatories. Both the ASCA and the {\it Beppo}SAX spectra required a moderately broadened Fe K$\alpha$ line (Reynolds \cite{reynolds}; Guainazzi et al. \cite{guai}).
\par
In Section 2 we discuss the observation and data reduction for both {\it Chandra} LETGS and XMM-{\it Newton}. The data analysis is described in Section 3. In Section 4 the luminosity variations for the LETGS data are analyzed, and in Section 5 we compare the LETGS spectrum with the RGS spectra. 

\section{Observations and data reduction}
\label{sec:obs}
The {\it Chandra} data were obtained on the 16$^{\rm th}$ of February 2001. NGC~4593 was observed in the standard configuration for the Low Energy Transmission Grating (LETG) in combination with the HRC-S camera. The total exposure time of the observation was 108000 s. The spectral data extraction procedure is identical to that described in Kaastra et al. (\cite{kaastra02}) for NGC~5548. 
\par
The XMM-{\it Newton} data were obtained on the 2$^{\rm nd}$ of July 2000. All instruments onboard XMM-{\it Newton} looked at the source with different exposure times, Table~\ref{tab:xmm} lists the exposure times used in the analysis. The EPIC cameras were shut down early due to a high radiation background. The main results from XMM-{\it Newton} in this paper are based on the Reflection Grating Spectrometer (RGS). Due to the high radiation background the spectrum from RGS has a strong background component, about 50 times higher than during a normal observation. As a result, the background count rates are between half and a third of the total count rate. As the high radiation background lasted for more than 70~\% of the observation, the entire observation of RGS was used in the analysis below. For RGS 1 all CCDs still functioned, while for RGS 2 CCD 4 did not function. The XMM-{\it Newton} EPIC,  RGS and OM data were reduced with the standard SAS software, version 5.3.3. This version includes the calibration of the instrumental oxygen edge and up-to-date effective areas for RGS. All plots are shown in the observed frame.
\begin{table}
\caption{The modes and exposure times used in the analysis for the different XMM-{\it Newton} instruments.}
\label{tab:xmm}
\begin{tabular}{|l|l|l|}\hline
Instrument  &  exposure (s) & observational mode   \\
RGS 1 and 2 &  27000        & spectroscopy         \\
EPIC MOS 2  &  8000         & small window with medium filter \\
EPIC pn     &  6000         & small window with thin filter  \\
OM          &  6000         & visible grism           \\
OM          &  4500         & UVW1 ($245-320$ nm)     \\
OM          &  10000        & UVW2 ($180-225$ nm)      \\\hline
\end{tabular}
\end{table}

The OM was operated in imaging mode with three consecutive filters: visible grism, UVW1 (245-320 nm) and UVW2 (180-225 nm). The source was not bright enough - and the straylight background was too high - for a satisfactory grism spectrum to be obtained. The UV source fluxes were obtained using the omsource task. A source extraction region 6'' in diameter was used, and the background was estimated from an annular region surrounding the source. The resulting fluxes ( Blustin et al. \cite{blustin}), corrected for coincidence loss, deadtime, background and Galactic reddening, were F$_\nu$ of 2.46 $\times$ 10$^{-14}$ and 1.58 $\times$ 10$^{-14}$ W m$^{-2}$ for UVW1 and UVW2 respectively. The errors on these fluxes are around 10 \%.

\begin{figure}
\resizebox{\hsize}{!}{\includegraphics[angle=0]{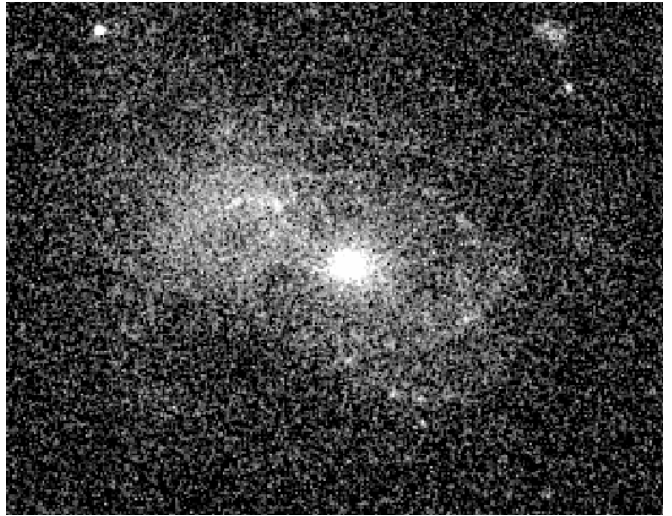}}
\caption{Image of NGC 4593 taken with the UVW1 filter of the Optical Monitor onboard of XMM-{\it Newton}. Besides the bright core the spiral arm structure is visible.}
\label{fig:om}
\end{figure}

Fig.~\ref{fig:om} gives the Optical Monitor image of NGC 4593 taken with the UVW1 filter. Other than the central core, some spiral arm structure can be seen from the image. The image taken with the UVW2 filter shows only a point source and has a rather low count rate in general. These images indicate that in the X-ray spectra there should be little contamination of the spiral arms, consistent with the point source detected in the EPIC instruments.

\section{Data analysis}
\label{sec:analyse}
\subsection{The warm absorber model}

\begin{figure}
\resizebox{\hsize}{!}{\includegraphics[angle=-90]{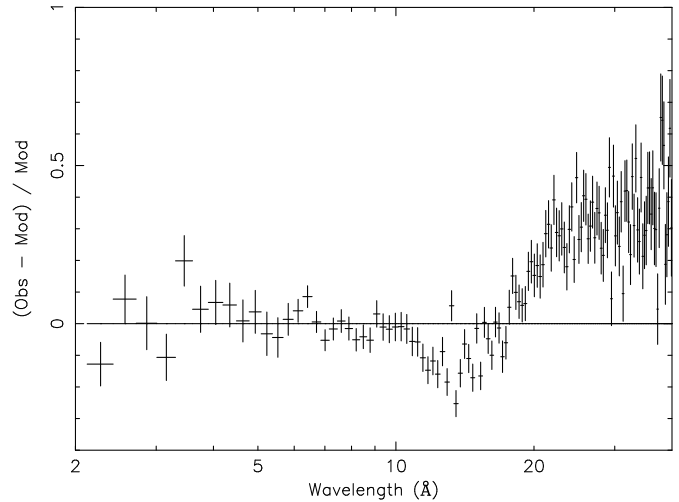}}
\caption{Fit residuals of a power-law fit to the LETGS data for NGC~4593, clearly showing a dip in the spectrum between 10 and 18 \AA~and the strong soft excess above 18\AA. The power-law was fit between 2 and 10 \AA~and the data are binned by a factor of 10.}
\label{fig:fig1}
\end{figure}

In this paper we use the {\it slab} and {\it xabs} model in SPEX (Kaastra, Mewe \& Nieuwenhuijzen \cite{kaastra96}; Kaastra et al. \cite{kaastrasp}) for the modeling of the warm absorber. The {\it slab} model calculates the transmission of each ion individually, using all absorption lines and edges. The transmission of all ions is then combined to calculate the total transmission of the warm absorber. All ions have the same outflow velocity and velocity broadening (Kaastra et al. \cite{kaastrasp}). 
The {\it xabs} model is the same as the slab model, except that the ion column densities are coupled by a grid of XSTAR photoionization models (Kallman \& Krolik \cite{kall}), characterized by the ionization parameter, $\xi$. For more details see Kaastra et al. (\cite{kaastrasp}). All quoted errors are 1$\sigma$ errors.
\par
The {\it xabs} model has the advantage that ions, which are too weak to be detected individually, are automatically included to give a consistent ionization parameter and column density. The drawback is that the {\it xabs} model is more dependent on the details of the photoionization model, than a simple ion by ion fit.
\par
For both models one can fit an overall blue- or redshift to the observed lines. Other parameters, which were not left free, as the  detected absorption lines are too weak, are the covering factor, the broadening due to a range in velocities, the width and separation of the velocity components (as obtained or estimated from, for example, UV data). The standard values for these parameters are as follows: covering factor is 1, velocity broadening is 250~km\,s$^{-1}$ and the width of the individual velocity components and their separation are both 100~km\,s$^{-1}$.

\subsection{Spectral analysis: {\it Chandra}}
The fluxed LETGS spectrum shows a dip between $10-18$ \AA~(see Fig.~\ref{fig:fig1}). However, few absorption lines can be detected by eye from the spectrum. The absorption edges of \ion{O}{vii} or \ion{O}{viii} are not detected, although a narrow \ion{O}{vii} forbidden and resonance line are observed at 22.3 and 21.8 \AA~(see Fig.~\ref{fig:fig4}). It is clear that no conventional warm absorber is detected. 
\par
\begin{figure}
\resizebox{\hsize}{!}{\includegraphics[angle=-90]{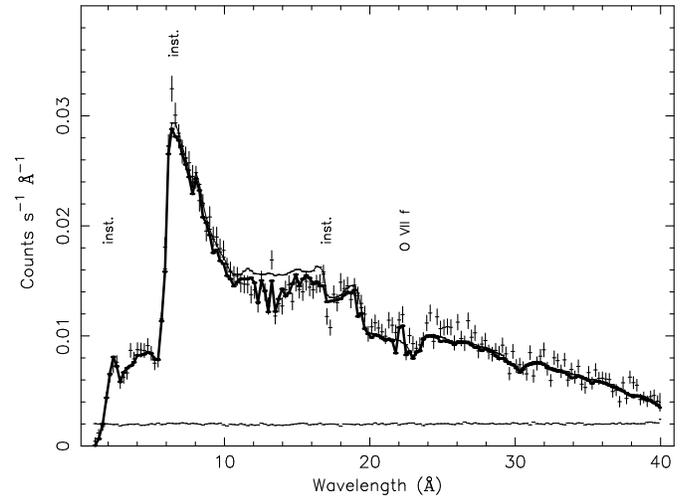}}
\caption{Power-law plus modified black body fit (thin solid line) and the model including the {\it xabs} components as described in Table~\ref{tab:xabsfit} (thick solid line) to the LETGS data. The thin line at about 0.002 counts s$^{-1}$ \AA$^{-1}$ is the subtracted background contribution. For clarity the data are binned by a factor of 5. Instrumental edges are labeled, as well as the \ion{O}{vii} forbidden line.}
\label{fig:fig2}
\end{figure}

Fitting the LETGS data with a power-law (PL) with galactic absorption gave a rather good fit to the data, except for the $10-18$ \AA~region, and at longer wavelengths where there is a soft excess (see Fig.~\ref{fig:fig1}). Adding a modified black body  (MBB) component (Kaastra \& Barr \cite{kaastrab}), does improve the fit at longer wavelengths, but does not explain the dip between $10-18$ \AA~(see Fig.~\ref{fig:fig2}). The dip in the spectrum cannot be attributed to calibration uncertainties or well fitted with another model for the soft excess, therefore we fitted the data including also a {\it xabs} component for the warm absorber (Kaastra et al. \cite{kaastrasp}).
\par
\begin{figure}
\resizebox{\hsize}{!}{\includegraphics[angle=-90]{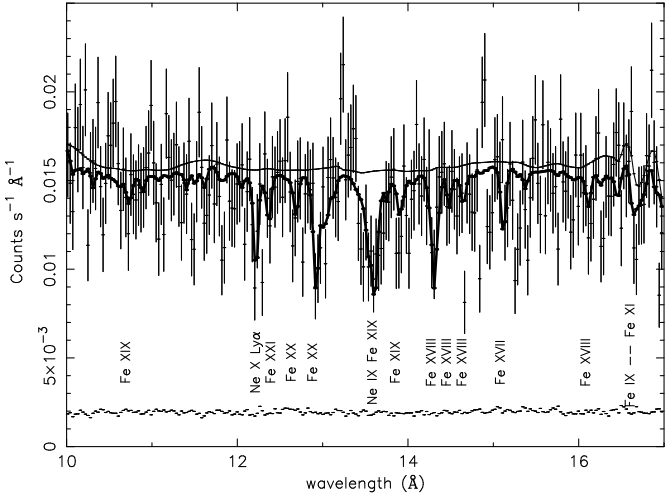}}
\caption{ Detailing the $10-17$ \AA~region of Fig.~\ref{fig:fig2}. Power-law and modified black body fit (thin solid line) and the fit including the {\it xabs} components (thick solid line) to the unbinned LETGS data.}
\label{fig:fig3}
\end{figure}
\begin{figure}
\resizebox{\hsize}{!}{\includegraphics[angle=-90]{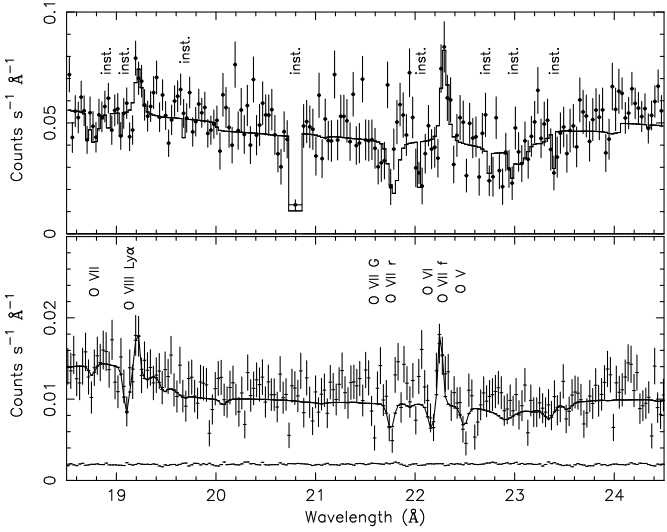}}
\caption{Detail of the RGS data (upper panel) and the LETGS data (lower panel). In the LETGS panel the absorption lines are indicated, G stands for Galactic absorption. In the upper panel the instrumental features for the RGS are labeled. The subtracted background is indicated for the LETGS data.}
\label{fig:fig4}
\end{figure}

 The results of the fits with the {\it xabs} component are summarized in Table~\ref{tab:xabsfit}. In Fig.~\ref{fig:fig2} and ~\ref{fig:fig3} the model with and without the {\it xabs} component are plotted, while Table~\ref{tab:chi} gives the $\chi$$^{2}$ and degrees of freedom for the different models.
\begin{table}
\caption{Fit parameters for the LETGS spectrum,
assuming a distance of 50.0 Mpc for NGC~4593.}
\label{tab:xabsfit}
\begin{tabular} {|l|l|c|} \hline
PL:&norm$^{a}$ & (2.18 $\pm$ 0.02) $\times$10$^{51}$ ph s$^{-1}$ keV$^{-1}$\\
           &$\Gamma$  & 1.69 $\pm$ 0.02 \\\hline
MBB:&norm$^{b}$ & (1.0 $\pm$ 0.1) $\times$10$^{32}$ m$^{1/2}$\\
            &T & (0.13 $\pm$ 0.01) keV \\\hline
\ion{O}{vii}f:&EW& ($-$152 $\pm$ 45) m\AA \\
       &flux& (0.45 $\pm$ 0.13) ph s$^{-1}$ m$^{-2}$ \\
       &$\lambda$$^{c}$& (22.069 $\pm$ 0.016) \AA \\
       &v$^{d}$& ($-$430 $\pm$ 218) km\,s$^{-1}$ \\\hline
{\it xabs} :&$N_{\rm H}$ & (1.6 $\pm$ 0.4) $\times$10$^{25}$ m$^{-2}$\\
      &log $\xi$$^{e}$ & 2.61 $\pm$ 0.09 \\
      &v$^{d}$& ($-$400 $\pm$ 121) km\,s$^{-1}$ \\
abun$^{f}$:&\ion{O} & 0.2 ($-$0.1, + 0.2)  \\\hline
{\it xabs} :&$N_{\rm H}$ & (6 $\pm$ 3) $\times$10$^{23}$ m$^{-2}$\\
      &log $\xi$$^{e}$ & 0.5 $\pm$ 0.3 \\
      &v$^{d}$& ($-$380 $\pm$ 137) km\,s$^{-1}$ \\\hline
\end{tabular}\\
             \\
$^{a}$ at 1 keV.\\
$^{b}$ norm=emitting area times square root of electron density.\\
$^{c}$ wavelengths are given in the rest frame of NGC~4593.\\
$^{d}$ velocity shift from comparing rest and observed wavelengths.\\ 
$^{e}$ in 10$^{-9}$ W m or in erg cm s$^{-1}$.\\
$^{f}$ only non-solar ratio abundances are noted; abundances relative \\
       to Fe, which is tied to solar abundance.
\end{table}

\begin{table}
\caption{The $\chi$$^{2}$ and degrees of freedom, as well as the significance of the added component according to an F-test for the different model fits to the LETGS data, as described in the text. Best fit parameters of model 6 are listed in Table 2. Z stands for abundances.}
\label{tab:chi}
\begin{center}
\begin{tabular} {ll|llll}
      &model          & $\chi$$^{2}$ & dof   & ${\chi_\nu}^2$ & sign.\\\hline
1     &PL             & 2073         & 1312  & 1.58   &       \\
2     &PL+MBB         & 1795         & 1310  & 1.37   & 0.99  \\
3     &PL+MBB+\ion{O}{vii}f& 1775    & 1308  & 1.36   & 0.58  \\
4     &3+{\it xabs}   & 1603         & 1305  & 1.23   & 0.97  \\
5     &4+Z free       & 1575         & 1298  & 1.21   & 0.62  \\
6     &5+1 {\it xabs} & 1556         & 1295  & 1.20   & 0.59  \\
\end{tabular}
\end{center}
\end{table}

\par
Using the {\it xabs} component and allowing only the ionization parameter $\xi$, the X-ray column density, $N_{\rm H}$ and the outflow velocity to be free parameters, we found a high ionization parameter of log $\xi$ = 2.6 in units of 10$^{-9}$ W m. This component describes the $10 - 18$ \AA~dip even if solar abundances (Anders \& Grevesse \cite{anders}) are assumed. There is an improvement in the fit if the abundances are left free. For this high ionization parameter Ar, Ne and Fe are the main elements absorbing the power-law component, as most other abundant elements such as C and N are already fully ionized. \ion{O}{viii} is the only oxygen ion expected in the spectrum. 
\par
Most of the Fe-absorption lines have small optical depths causing rather shallow lines. These are, therefore, not detectable per individual line, with the current sensitivity. The combination of these lines causes a significant depression in the continuum spectrum and blending results in some observable broadband absorption structures (see Fig.~\ref{fig:fig3}). The abundance quoted in Table~\ref{tab:xabsfit} is effectively measured relative to Fe, which was fixed to solar abundance, because of its many absorption lines. Compared to iron, oxygen is underabundant, while all other elements have abundance ratios to iron consistent with the solar ratio. 
\par
A weak, redshifted emission component for the \ion{O}{viii} Ly$\alpha$ line is observed (see Fig.~\ref{fig:fig4}), which is partially blended with the absorption component. However, the non-detection of an \ion{O}{viii} Ly$\gamma$ absorption line, and the weak \ion{O}{viii} Ly$\beta$ constrain the column density of \ion{O}{viii} and thus also the abundance ratio for oxygen. The oxygen abundance in Anders \& Grevesse (\cite{anders}) might be too high (see Allende Prieto, Lambert \& Asplund \cite{allende}) resulting in the derived underabundance.
\par
From a more detailed inspection of the LETGS spectrum we also observe weak absorption lines from \ion{O}{v} as well as \ion{O}{vi} in the spectrum (see Fig.~\ref{fig:fig4}). These lines represent a low ionization component and to account for them we added a second {\it xabs} component to our model. The ionization parameter for this second component is log $\xi$ = 0.5 in units of 10$^{-9}$ W m. However, the total column density for this component is about 25 times smaller than for the high ionization component (see Table~\ref{tab:xabsfit}).

\subsection{Spectral analysis: RGS}

The RGS data, taken with the XMM-{\it Newton} satellite are noisy due to the short exposure time and the high background radiation level. As a result, it is hard to constrain any model for the warm absorber. We fitted RGS 1 and 2 simultaneously, but for clarity added RGS 1 and 2 in the figures. The only significant narrow features in the spectrum are the \ion{O}{vii} forbidden line in emission (see Fig.~\ref{fig:fig4}), and the \ion{Ne}{ix} resonance line in absorption. We applied the model used for the analysis of the {\it Chandra} data: a power-law, modified black body, galactic absorption, the forbidden \ion{O}{vii} emission line and a {\it xabs} component. Due to the noise in the data, however, the {\it xabs} component is hard to constrain. Therefore we fitted the warm absorber with the {\it slab} component (see Table~\ref{tab:rgsslab}). From the {\it slab} model we find a rather large spread in ionization parameters. 
\par
The {\it xabs} model could not constrain the absorption, because not all ions at a certain ionization state are observed. For instance, \ion{O}{viii} and \ion{Ne}{x} are too weak to be detected, while highly ionized iron is observed. A possible explanation is that the \ion{O}{viii} Ly$\alpha$ absorption line is partly blended by its emission component. A narrow \ion{O}{viii} Ly$\alpha$ line in emission is observed with 1.6 $\sigma$ significance. This absorption model is consistent with the higher S/N LETGS data.
\par
Interestingly, in the RGS data set we see evidence for galactic absorption from \ion{O}{vii} and \ion{O}{viii}, with logarithms of the column densities, in m$^{-2}$, of 20.2 $\pm$ 0.5 and 20.1 $\pm$ 0.9 respectively. For the LETGS data  the galactic \ion{O}{vii} resonance line, which is the most prominent galactic line, has an equivalent width (EW) of 45 $\pm$ 31 m\AA.
\begin{table}
\caption{The different $\chi$$^{2}$ and degrees of freedom, as well as the significance of the added component according to an F-test for the different models fit to the RGS data, as described in the text.}
\label{tab:rgschi}
\begin{center}
\begin{tabular}{ll|llll}
      &model          & $\chi$$^{2}$     & dof  & ${\chi_\nu}^2$ & sign. \\\hline
1     &PL+MBB         & 1317             & 1045 & 1.26 &         \\
2     &1+\ion{O}{vii}f& 1298             & 1043 & 1.24 & 0.59    \\
3     &2+\ion{Ne}{ix}r& 1275             & 1041 & 1.22 & 0.61    \\
4     &1+{\it xabs}   & 1260             & 1040 & 1.21 & 0.76    \\
5     &1+{\it slab}   & 1186             & 1020 & 1.16 & 0.95    \\
6     &5+rel.\ion{N}{vii} Ly$\alpha$&1166& 1009 & 1.16 & 0.61    \\\hline
7     &2+3rel.lines   & 1162             & 1035 & 1.12 & 0.96    \\
\end{tabular} 
\end{center}
\end{table}

\begin{table}
\caption{Continuum and emission parameters, describing the preferred model (i.e model 6 in Table~\ref{tab:rgschi}), for RGS1 and RGS2, assuming a distance of 50.0 Mpc for NGC~4593.}
\label{tab:rgsfit}
\begin{tabular} {|l|l|c|} \hline
PL:&norm$^{a}$ & (3.86 $\pm$ 0.2) $\times$10$^{51}$ ph s$^{-1}$ keV$^{-1}$\\
   &$\Gamma$   & 1.8 $\pm$ 0.2 \\
MBB:&norm$^{b}$& (1.4 $\pm$ 0.4) $\times$10$^{32}$ m$^{1/2}$ \\
    & T        &  (0.16 $\pm$ 0.01) keV \\       
\ion{O}{vii}f:&EW& ($-$110 $\pm$ 34) m\AA\\
       &flux& (0.84 $\pm$ 0.28) ph s$^{-1}$ m$^{-2}$ \\
       &$\lambda$$^{c}$ & (22.12 $\pm$ 0.02) \AA \\\hline
\multicolumn{3}{c}{Relativistic emission line:}\\\hline
       & i (degrees)  & 30 $\pm$ 11 \\
       & q $^{d}$     & 3 $\pm$ 6  \\
       & $R_{\rm in}$ ($GM/c^2$) & $<$ 89 \\
       & $R_{\rm out}$ ($GM/c^2$) & 400  \\
\ion{N}{vii}: &norm& (2.1 $\pm$ 0.9) ph s$^{-1}$ m$^{-2}$\\
       &EW            & (0.3 $\pm$ 0.2) \AA \\
       &$\lambda$$^{c}$& (24.9 $\pm$ 0.8) \AA\\\hline
\end{tabular}\\
             \\
$^{a}$ at 1 keV.\\
$^{b}$ norm = emitting area times square root of electron density.\\
$^{c}$ wavelengths are given in the rest frame of NGC~4593.\\
$^{d}$ the emissivity slope.
\end{table}

\begin{table}
\caption{The logarithms of the column density in m$^{-2}$ for absorption in the NGC~4593 RGS data.}
\label{tab:rgsslab}
\begin{center}
\begin{tabular} {|lr|lr|}
\hline 
\ion{C}{vi} & 20.3 $\pm$ 0.4  & \ion{Fe}{viii} & 20.0 $\pm$ 0.7  \\
\ion{N}{vi} & 20.0 $\pm$ 0.4  & \ion{Fe}{ix}   & 20.1 $\pm$ 0.4 \\
\ion{O}{vii}& 21.1 $\pm$ 0.5  & \ion{Fe}{xiii} & 19.8 $\pm$ 0.4 \\
\ion{Ne}{ix}& 20.6 $\pm$ 0.5  & \ion{Fe}{xvi}  & 19.9 $\pm$ 0.8 \\
\ion{Na}{xi}& 21.1 $\pm$ 0.7  & \ion{Fe}{xvii} & 19.9 $\pm$ 0.5 \\
\ion{Si}{ix}& 21.0 $\pm$ 0.5  & \ion{Fe}{xviii}& 20.1 $\pm$ 0.4 \\
\ion{Si}{xii}&20.7 $\pm$ 0.6  & \ion{Fe}{xix}  & 20.5 $\pm$ 0.6 \\
\ion{S}{xii}& 19.7 $\pm$ 0.7  & \ion{Fe}{xx}   & 20.5 $\pm$ 0.3 \\
\ion{Ar}{x} & 19.6 $\pm$ 0.6  & \ion{Fe}{xxi}  & 20.5 $\pm$ 0.3 \\
\ion{Ar}{xii}&19.5 $\pm$ 0.7  &                &                \\\hline
\end{tabular}\\
\end{center}
\end{table}

\begin{figure}
\resizebox{\hsize}{!}{\includegraphics[angle=-90]{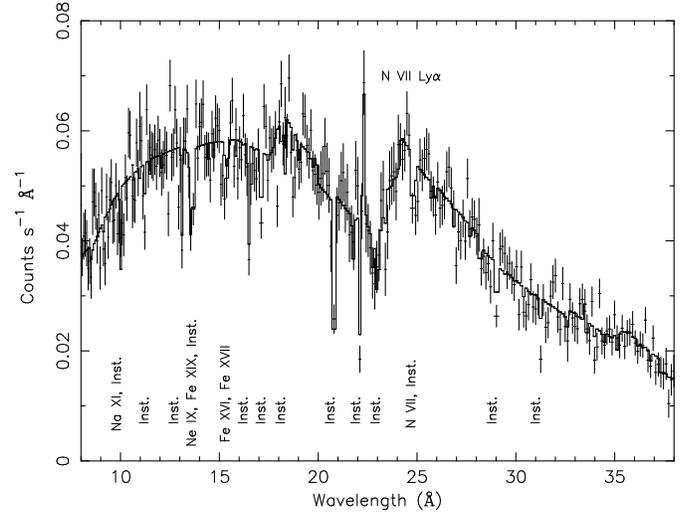}}
\caption{Fit to the RGS data with model 6 of Table~\ref{tab:rgschi}. For clarity RGS 1 and RGS 2 were binned by a factor of 10 and added. The relativistic \ion{N}{vii} Ly$\alpha$ line is indicated by \ion{N}{vii} Ly$\alpha$ written horizontally. The features labeled Inst. are periodic dips due to CCD-gaps, or due to instrumental calibration.}
\label{fig:fig5}
\end{figure}

\begin{figure}
\resizebox{\hsize}{!}{\includegraphics[angle=-90]{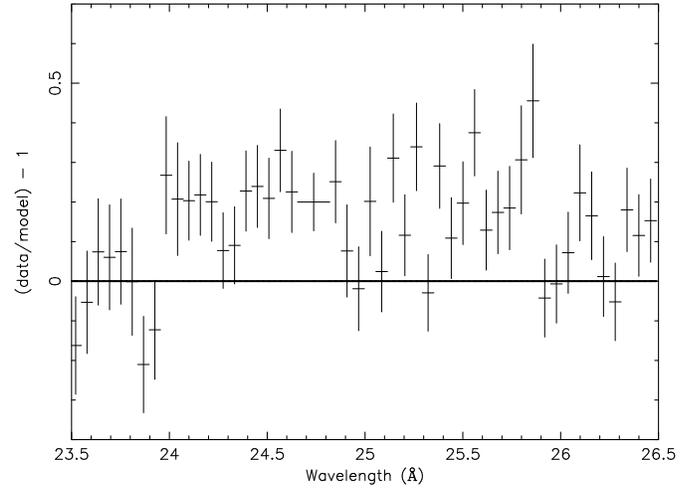}}
\caption{Fit residuals between 23.5 and 26.5 \AA, showing the observed excess at the \ion{N}{vii} Ly$\alpha$ wavelength, which is fitted in model 6 (Table~\ref{tab:rgschi}) with a relativistic emission line.}
\label{fig:fign7}
\end{figure}

An excess above the continuum is noted at 24.9 \AA~(see Fig.~\ref{fig:fign7}). This excess is consistent with an extremely broadened relativistic emission line from \ion{N}{vii} Ly$\alpha$, and the normalization has a 2 $\sigma$ significance (see Table~\ref{tab:rgsfit}). The disk inner radius implied by this emission line corresponds to $\sim$0.6 $c$. We found no significant \ion{O}{viii} and \ion{C}{vi} Ly$\alpha$ lines. The model including the \ion{N}{vii} relativistic line and the warm absorber (i.e. model 6 in Table~\ref{tab:rgschi}) is plotted in Fig.~\ref{fig:fig5}.

As a second model, we fitted the RGS data with only a power-law, modified black body and three relativistic emission lines, namely, for \ion{O}{viii}, \ion{N}{vii} and \ion{C}{vi} Ly$\alpha$. This model (i.e. model 7 in Table~\ref{tab:rgschi}) has a $\chi^{2}$ of 1162 for 1035 degrees of freedom, a rather flat photon index of 1.4, and a 30 \% lower normalization. The modified black body parameters are not as sensitive, and consistent within 3 $\sigma$ of those quoted in Table~\ref{tab:rgsfit}. In this model we find a 3 $\sigma$ detection for the \ion{N}{vii} Ly$\alpha$ and a 2 $\sigma$ measurement for the \ion{O}{viii} Ly$\alpha$ line. The \ion{C}{vi} Ly$\alpha$ line is not detected in both models. In this model only absorption by \ion{Ne}{ix} is detected with a 3 $\sigma$ significance. However, this model over predicts the flux at higher energies, compared to those measured by pn and MOS2. This is due to the flat photon index. As a result we exclude this model, and prefer the absorption model (model 6 in Table~\ref{tab:rgschi}) as the continuum is consistent with the one observed with the EPIC instruments. This model does include a relativistic \ion{N}{vii} Ly$\alpha$ line which was not detected in the LETGS data. In the {\it Chandra} LETGS spectrum of NGC~5548 (Kaastra et al. \cite{kaastra02}) the observed equivalent width for the relativistic \ion{N}{vii} Ly$\alpha$ is nearly twice that of \ion{O}{viii} Ly$\alpha$. Our findings are thus consistent with some earlier results on relativistic emission lines. The absence of the relativistic \ion{N}{vii} Ly$\alpha$  line in the LETGS spectrum could indicate that the strength of this emission line is variable. Previous studies of relativistic lines in the soft X-ray band have shown some evidence for time variability (Steenbrugge et al. \cite{steen}). Due to the weakness of these relativistic lines, the EPIC data cannot be used to constrain either model further.

\subsection{EPIC continuum and Fe K$\alpha$}
The EPIC continuum was fitted with a power-law (PL), reflection (Refl) component and a modified black body (MBB). The power-law and reflection component were fitted in the $2 - 10$ keV range. Afterward we fitted a modified black body assuming the warm absorber parameters as determined from the RGS data (see Table~\ref{tab:rgsslab}), including the $0.3-2$ keV range. Note that for MOS the photon index for the reflection component is not given, as we could not constrain it. The differences between the best fit model with a warm absorber (continuum parameters given in Table~\ref{tab:mosfit}) and without are negligible, certainly for the MOS data. In both datasets the models overpredict the count rate between 0.8 and 1 keV, even if the warm absorber is included.

\begin{figure}
\resizebox{\hsize}{!}{\includegraphics[angle=-90]{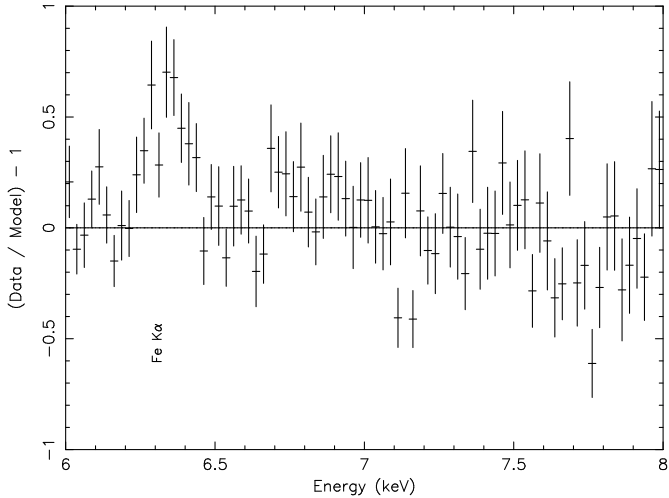}}
\caption{The relative fit residuals with respect to a simple power-law fit for the pn and MOS2 data added together. The power-law was fit between 2 and 10 keV.}
\label{fig:fig6}
\end{figure}

The MOS and pn spectra show clear evidence for a narrow Fe K$\alpha$ line (see Fig.~\ref{fig:fig6}), but no broad component can be detected. The line was modeled with a Gaussian profile and is detected with a 3 $\sigma$ significance in both datasets. The fit results for pn and MOS2 are given in Table~\ref{tab:mosfit}. No relativistically broadened Fe K$\alpha$ component has been found before in NGC~4593.  
\begin{table}
\caption{pn and MOS2 results for the continuum and the \ion{Fe} K$\alpha$ line.}
\label{tab:mosfit}
\begin{tabular} {|l|c|c|} \hline
                            & pn              & MOS2            \\\hline
PL norm$^{a}$               & 3.3 $\pm$ 0.1   & 2.9 $\pm$ 0.2   \\
$\Gamma$                    & 1.83 $\pm$ 0.02 & 1.96 $\pm$ 0.05 \\
Refl norm$^{a}$             & $<$ 0.7         & $<$ 0.5         \\
Refl $\Gamma$               & 1.97 $\pm$ 0.14 &                 \\
MBB norm$^{b}$              & 1.5 $\pm$ 1     & 8.8 $\pm$ 6     \\
T (keV)                     & 0.16 $\pm$ 0.005& 0.09 $\pm$ 0.008 \\\hline
flux Fe K$\alpha$$^{c}$     & 0.42 $\pm$ 0.09 & 0.37 $\pm$ 0.12 \\
FWHM (keV)                  & $<$ 0.13        & $<$ 0.55        \\
rest E (keV)                & 6.40 $\pm$ 0.05 & 6.40 $\pm$ 0.05 \\\hline 
\end{tabular}\\
              \\
$^{a}$ in 10$^{51}$ ph s$^{-1}$ kev$^{-1}$ \\
$^{b}$ in 10$^{32}$ m$^{1/2}$ \\
$^{c}$ ph s$^{-1}$ m$^{-2}$
\end{table}
\section{Timing analysis}
\begin{figure}[ht]
\resizebox{\hsize}{!}{\includegraphics[angle=-90]{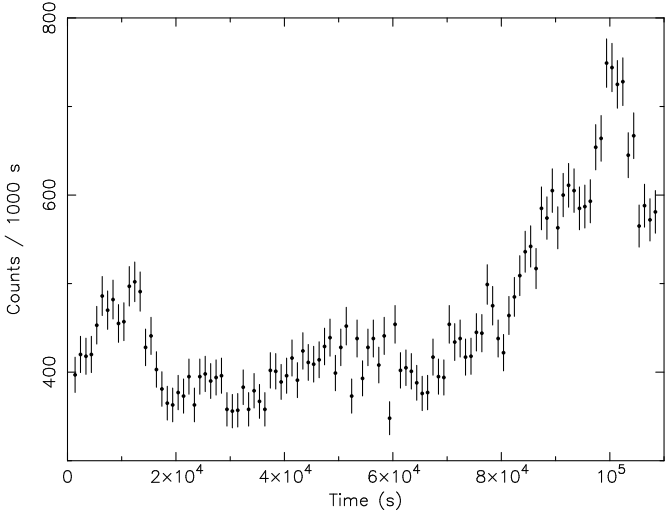}}
\caption{The light curve for the LETGS data, in the zeroth spectral order. Note the rise phase between 70 and 97 ks, and the peak phase between 97 and 105 ks.}
\label{fig:fig7}
\end{figure}

In the long {\it Chandra} LETGS observation, we observe a flare near the end of the observation. At the peak of the flare the luminosity increases by a factor of $\sim$1.5 in about 27 ks (see Fig.~\ref{fig:fig7}). The flare peak lasted about 7 ks, afterward there is a 15 ks period were the flux level decreased and possibly leveled off. Due to the low count rates measured and the relatively short duration of the peak, only a broadband comparison between the peak and the quiescence state is possible. 
\par
\begin{figure}
\resizebox{\hsize}{!}{\includegraphics[angle=-90]{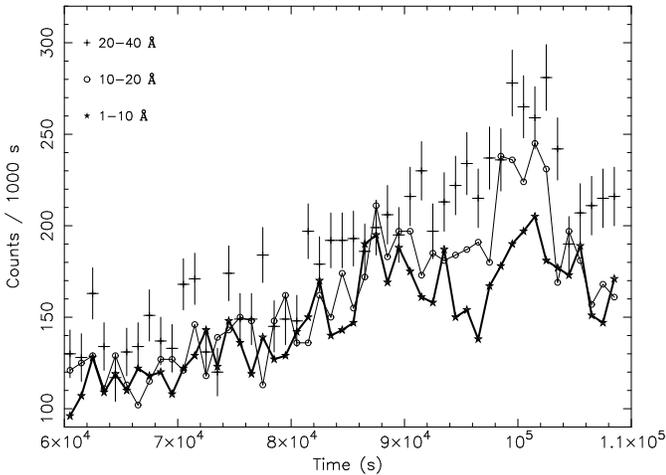}}
\caption{The light curve during the rise and peak phase for the LETGS data, but now split up into three wavelength bands. The stars joined by the thick line represent the power-law component, or the 1-10 \AA~photons. The crosses represent the soft excess component, i.e. the 20-40~\AA~photons and the open circles joined by a thin line are the intermediate wavelength range of 10-20 \AA. The errors for the power-law and intermediate component are similar to those plotted for the soft excess component.}
\label{fig:fig8}
\end{figure}

We took the spectrum of quiescence and the peak separately and binned them to 1 \AA~bins in order to have reasonable errors for the data points. The flux increase during the peak was mainly wavelength independent above 15 \AA, while the flux increase at lower wavelengths was significantly smaller (Kaastra \& Steenbrugge \cite{kaastra01}). We separate the light curve into three components: one representing the power-law component ($1-10$ \AA), one representing the soft excess component ($20-40$ \AA) and finally an intermediate component between $10-20$ \AA. A detail of the lightcurves, indicating the softening of the spectrum during the rise and peak phase is shown in Fig.~\ref{fig:fig8}. Also during the smaller flare in the beginning of the observation the spectrum is softer than during quiescence (between 20 ks and 70 ks). 
\par
The major contribution to the flare was an increase in the soft excess, i.e. from the accretion disk spectrum. The 10 $-$ 20~\AA~component is near constant during the rise phase with respect to the 1 $-$ 10 \AA~component, but does rise during the peak. The rise of the power-law component is smaller and more complex: it has two maxima (83 and 102 ks) while the soft excess rises smoothly to the flare peak (87 and 102 ks). This is not consistent with the picture that the power-law is produced solely by Inverse Compton scattering of UV and soft X-ray photons. Several authors have noted that the power-law component is variable on time scales smaller than the variability detected in the soft excess (see e.g. Dewangan et al. \cite{dewan}; Turner et al. \cite{turner}). A possible explanation is that magnetic flares take place in a corona (Merloni \& Fabian \cite{merloni}) on these shorter time scales. 
\par
The power-law component decays, after the first peak, in 9~ks between 87.5 ks and 96.5 ks. We derive a half-life of 18 ks. This decay seems independent of the soft excess component, and might thus be representative of a magnetic reconnection decay time. After the second peak also the soft excess component decays, therefore the decay of the power-law component can be caused by both magnetic reconnection decay and a decrease of seed photons for the Inverse Compton scattering.
\par
For the XMM-{\it Newton} data the study of luminosity variations is complicated due to the short observation time and the high radiation background. Over the good time interval, of only 8000 s, the luminosity was constant.


\section{Discussion}
The 2-10 keV luminosity of NGC~4593 during the XMM-{\it Newton} observation was 1.2$\times$10$^{36}$ W, estimated from model 6 in Table~\ref{tab:rgschi}. For pn and MOS2 we find 1.1$\times$10$^{36}$ W and 8.5$\times$10$^{35}$ W respectively. The LETGS luminosity was 9.0$\times$10$^{35}$ W and during the ASCA observation it was 1.15$\times$10$^{36}$ W (Reynolds \cite{reynolds}). The temperatures found for the MBB are consistent at the 3 $\sigma$ level between the {\it Chandra} LETGS and XMM-{\it Newton} results, while the photon index is consistent between the LETGS and RGS data, but inconsistent at the 3 $\sigma$ level between the LETGS and EPIC results.
\par
Guainazzi et al. (\cite{guai}) found a moderately broad neutral Fe K$\alpha$ line, with $\sigma$ $>$ 60 eV or a FWHM greater than 141 eV. Reynolds (\cite{reynolds}) quotes a $\sigma$ of 66 $\pm$ 22 eV or equivalently a FWHM of 155 $\pm$ 52 eV. The upper limit to the FWHM for the Fe K$\alpha$ line from the pn instrument is 130 eV. This is consistent with the result from Reynolds (\cite{reynolds}), but is marginally inconsistent with the best fit results obtained by Guainazzi et al. (\cite{guai}) from the {\it Beppo}SAX data. The results from MOS2 are less constraining and consistent with both previous observations.
\par
Comparing our spectra with the earlier {\it Beppo}SAX observation, we find a similar power-law slope for our XMM-{\it Newton} data set. Guainazzi et al. (\cite{guai}) note an excess between 0.3 and 0.6 keV in the {\it Beppo}SAX spectrum, but conclude it is most likely a calibration feature and not a soft excess. However, we find a significant soft excess in both the {\it Chandra} LETGS and XMM-{\it Newton} observations. Guainazzi et al. (\cite{guai}) explained the dip around 0.7 keV as due to absorption edges of \ion{O}{vii}, \ion{O}{viii}, \ion{Ne}{ix} and \ion{Ne}{x}. For the LETGS spectrum we explain this dip by absorption of highly ionized iron and neon ions. Also in the RGS there is evidence for absorption from highly ionized ions. We cannot compare the reflection component detected in the {\it Beppo}SAX data, as our dataset cuts off at around 10 keV, and the reflection component is minimal there. The width of the FeK$\alpha$ line derived from the EPIC data is consistent with those derived from the {\it Beppo}SAX and ASCA observations (Guainazzi et al. \cite{guai}; Reynolds \cite{reynolds}). However, in a more recent observation Yaqoob \& Padmanabhan (\cite{yaqoob}) conclude from the line intensity difference between simultaneous {\it Chandra} HEG and RXTE PCA observations that there is a broad component to the Fe K$\alpha$ line.
\par
Both the LETGS and the RGS data show a prominent \ion{O}{vii} forbidden emission line. The fluxes measured for this line in the two observations are consistent and the line is unresolved in both cases. We expect that this narrow emission line is formed further out from the ionization source than the absorption lines.
\par
For the {\it Chandra} LETGS observation we observe two distinct warm absorbers which have an ionization state different by two orders of magnitude. The high ionization component is only detected through shallow, highly ionized iron and neon lines. The low ionization component has a 25 times smaller column density, but is represented by a few well detected absorption lines. From simple outflowing wind models a more continuous ionization range would be expected. A possible explanation is that the highly ionized warm absorber is only formed during the peak. However, the statistics are too poor to verify this. Due to the large errors the RGS data set is consistent with the absorber model derived from the LETGS spectrum. 

\section*{ACKNOWLEDGMENTS}
 
This work is based on observations obtained with XMM-{\it Newton}, an ESA science mission with instruments and contributions directly funded by ESA Member States and the USA (NASA). SRON National Institute for Space Research is supported financially by NWO, the Netherlands Organization for Scientific Research. The MSSL authors acknowledge the support of PPARC. EB was supported by the Yigal-Alon Fellowship and by the GIF Foundation under grant \#2028-1093.7/2001. KCS expresses gratitude for the hospitality of the Columbia Astrophysics Laboratory during her two months visit. The authors thank Chris Saunders for his contribution to early analysis of the XMM-{\it Newton} data.

\end{document}